# REFINEMENT OF STELLAR PARAMETERS FOR THE ECLIPSING BINARY SYSTEM KIC 8569819 USING STELLAR MODELING APPROACH

*J. A. D. M. Dharmathilaka[1,2*], J. Adassuriya[2], K. P. S. C. Jayaratne[2] and Jordi L. Gutiérrez[3]*

[1]*Department of Physical Sciences and Technology, Faculty of Applied Sciences, Sabaragamuwa University of Sri Lanka, Sri Lanka*

[2]*Astronomy and Space Science Unit, Department of Physics, Faculty of Science, University of Colombo, Sri Lanka*

[3]*Department of Physics, Universitat Politècnica de Catalunya (UPC), Spain*

Eclipsing binary systems with a Delta (δ) Scuti component serve a vital role in deriving precise fundamental stellar parameters and testing stellar evolution models. This study mainly focuses on the Kepler target KIC 8569819, a detached eclipsing binary system that consists of a δ Scuti pulsating component. The quarter 9 photometric data observed by the Kepler mission were used for the analysis. The binary nature of the KIC 8569819 system was modeled using the Wilson-Devinney (WD) code and extracted new set of stellar parameters. This comprehensive study mainly focuses on the application of the Differential Correction (DC2015) process after the initial fitting done by the Light Curve modeling (LC2015) process for the disentanglement of the binary nature from the observed light curve. Subsequently, an improved set of stellar parameters for both primary and secondary components of the KIC 8569819 system was determined. The DC2015 modeling process yielded an orbital inclination of i = 89.88 ± 0.03 degrees, primary component luminosity L = 10.911 ± 0.005 $L_\odot$, the effective temperature of the primary component of $T_{eff,1}$ = 7155 ± 9 K and the effective temperature of the secondary component of $T_{eff,2}$ = 5956 ± 7 K. Additionally, the values for the radius 1.790 $R_\odot$ and 0.986 $R_\odot$, bolometric magnitude 2.56 mag and 4.65 mag, and surface gravity 4.17 cm s$^{-2}$ and 4.46 cm s$^{-2}$, were found as refined stellar parameters for both primary and secondary components of the KIC 8569819 binary system respectively. These results not only deliver an updated and highly accurate stellar model for KIC 8569819 but also provide reliable input for the future analysis of mode identification of pulsation frequencies in the field of Asteroseismology.

[*]*Corresponding Author: dinesha@appsc.sab.ac.lk*

# REFINEMENT OF STELLAR PARAMETERS FOR THE ECLIPSING BINARY SYSTEM KIC 8569819 USING STELLAR MODELING APPROACH

*J. A. D. M. Dharmathilaka[1,2,\*], J. Adassuriya[2], K. P. S. C. Jayaratne[2] and Jordi L. Gutiérrez[3]*

[1]*Department of Physical Sciences and Technology, Faculty of Applied Sciences, Sabaragamuwa University of Sri Lanka, Sri Lanka*
[2]*Astronomy and Space Science Unit, Department of Physics, Faculty of Science, University of Colombo, Sri Lanka*
[3]*Department of Physics, Universitat Politècnica de Catalunya (UPC), Spain*

## INTRODUCTION

The study of stellar interior with the help of the oscillation pattern of stars is known as Asteroseismology (Aerts et al., 2010). Delta (δ) Scuti stars are the most important type of pulsating variables, which change their brightness due to periodic expansion and contraction of the surface layers of stars. The pulsation of the star can be radial or non-radial. They are located in the classical instability strip of the Hertzsprung-Russell (HR) diagram and generally their masses in between 1.4 and 2.5 solar masses ($M_\odot$) (Liakos & Niarchos, 2016). These stars exhibit two types of pulsation modes, namely, pressure mode (p-mode) and gravity mode (g-mode). p-mode is due to acoustic waves of the stars and pressure is the restoring force of the pulsation. The g-mode is due to the gravity waves of the stars and buoyancy is the restoring force of the pulsation. p modes are sensitive to the outer part of the stars, while g modes are sensitive to the inner part of the stars (Aerts et al., 2010).

A binary star system is a two-star system that is gravitationally bounding one star to another and revolves around a common center of mass. A particular subclass of binary star systems is eclipsing binaries. The pair of stars in the Eclipsing binaries are orbiting around each other, with the concept of light of one star being blocked by the other star according to the viewpoint of the observer (Liakos & Niarchos, 2016). If one star is a pulsating variable like δ Scuti, then the system is worth analyzing in terms of binary and pulsation characteristics. These systems can be subcategorized as detached binaries, semi-detected binaries and overcontact binaries (Liakos & Niarchos, 2016).

Wilson Devinney (pyWD2015) eclipsing binary modeling program is a widely used, valuable code for light curve modeling, stellar parameter refinement process and successful disentanglement of the binary nature from the observed light curves. The process of disentanglement of the binary nature of the eclipsing binary systems must be carried out more precisely to obtain highly accurate results for the use of frequency analysis and asteroseismic studies. The pyWD2015 program consists of two main processes, namely, light curve modeling or LC2015 and Differential Correction or DC2015. After fitting the model for the observed light curve through the LC2015 as the initial step, the DC2015 process must be run as the second step, for further refinement of the stellar parameters to increase the accuracy of the results generated by LC2015 (Güzel & Özdarcan, 2020; Kallrath, 2022; Wilson R. E., 2008; Wilson & Devinney, 1971). A general method of computing synthetic light curves is generated by integrating rotational and tidal distortion, the reflection effect, limb darkening and gravity darkening (Wilson & Devinney, 1971). Based on the study of Wilson & Devinney (1971), the WD program models binary systems under Roche geometry, where the stellar surfaces are defined by equipotential of dimensionless potential,

$$\Omega = r^{-1} + q\left[(1 - 2\lambda r + r^2)^{-\frac{1}{2}} - \lambda r\right] + \frac{1}{2}(q + 1)r^2(1 - \nu^2), \quad (1)$$

where, $q = m_2/m_1$ is the mass ratio, $\lambda$ and $\nu$ are direction cosines, $\Omega$ is a dimensionless potential function and the surface radius $r(\theta, \phi)$ at each direction is obtained by solving equation 1 (Wilson & Devinney, 1971).

Precise study for the light curve modeling and accurate results of the stellar parameters of the δ Scuti component in eclipsing binary system is required for the better understanding of the stellar interior of the pulsating component. The main focus of this study is to identify the best model for the light curve of KIC 8569819, which is a detached eclipsing binary system with δ Scuti component. Initially, the binary component was disentangled using the LC2015 method (Güzel & Özdarcan, 2020). Although prior studies of KIC 8569819 have modeled the light curve, this study focused on the DC2015 process for the first time for this target star. By enhancing the accuracy of the resultant basic stellar parameter values and yielding best disentanglement of binary nature from the observed light curve, the stellar parameters were refined.

**OBSERVATIONS**

Photometric data of Quarter (Q) 9 for the KIC 8569819 system were retrieved using the lightkurve package from the Kepler mission archive. The orbital period of the target eclipsing binary is approximately $20.84993 \pm 0.00003$ days (Liakos & Niarchos, 2016) and the photometry data relevant to the Q9 covers mainly 78 days of the observation period and which includes around three complete orbital cycles.

The initial set of stellar parameters for this system was adopted from Kurtz et al. (2015). In addition to the light curve modeling done by Kurtz et al. (2015), this study mainly revisited running the WD DC2015 program based on the Kepler photometry to improve the accuracy of the stellar parameters.

**BINARY LIGHT CURVE MODELING PROCESS BY WILSON DEVINNEY LC2015 PROGRAM**

The retrieved data from the Q9 of the Kepler mission were used for the process of cleaning outliers and then the system data were normalized to their median value for the use of the next process of the modelling.

The initial input parameters of the KIC 8569819 target, orbital period ($P_{orb}$), inclination angle (i), semi-major axis (a), modified potentials of primary star ($\Omega_1$) and secondary star ($\Omega_2$), mass ratio (q), luminosity of the primary star ($L_1$), effective temperature of primary star ($T_{eff,1}$) and secondary star ($T_{eff,2}$), limb darkening coefficients for the primary star ($x_1$) and secondary star ($x_2$) were obtained from the study of Kurtz et al. (2015). Based on the concept of whether a star's envelope is radiative or convective, the necessary input parameters of the gravity darkening coefficient (g) and albedo coefficient (A) were identified for the modeling process. For KIC 8569819 target, g and A values were assigned for the primary star as $g_1 = 1$ and $A_1 = 1$ while for the secondary star as $g_2 = 0.32$ and $A_2 = 0.6$. The epoch ($T_{epoch}$) is the reference time of our data set at the primary eclipse and it was calculated based on equation (2).

$$T_{epoch} = T_0 + (P_{orb} \times E) , \qquad (2)$$

where, $T_0$ is the initial time of the minimum light, and E is the number of orbital cycles (Darwish et al., 2024).

Among those input parameters, $T_{eff,1}$, a, $P_{orb}$, $\Omega_1$, $\Omega_2$, A, g, $x_1$ and $x_2$ were kept fixed during the modeling, while other parameters can be assigned as adjustable or fixed as required by the LC2015 modeling process (Güzel & Özdarcan, 2020; Kallrath, 2022). By following this whole process for the LC Program, we were able to match the observed light curve with the model and able to obtain the basic stellar parameters of mass and radius of the primary and secondary components as a result of the LC2015 program. Subsequently, the refinement process of stellar parameters was applied through the DC2015 process to generate a new set of stellar parameters for the star KIC 8569819.

## STELLAR PARAMETER REFINEMENT PROCESS BY WILSON DEVINNEY DC2015 PROGRAM AND RESULTS

In the DC2015 process, the output of the light curve model generated from the LC2015 was used for the refinement process of the stellar parameters. In the DC2015 process, the model is iteratively refined to match the observed light curves by reducing the deviation between observed and the modeled light curves (Güzel & Özdarcan, 2020; Kallrath, 2022; Wilson R. E., 2008; Wilson & Devinney, 1971). For the DC2015 process, we need to identify the external iteration number. It can be determined by checking the convergence criteria. If the convergence criteria are achieved for the KIC 8569819 for some value of external iteration, the green color flag appears in the 'output' label in the 'Results' tab. It says further improvement to the generated model and further refinement process of stellar parameters are not applicable. For this study, the best value for the external iteration was 6.

From that, we were able to obtain the final best model generated from the DC2015 program, as shown in Figure 1, while the phased light curve of the system with its final fit is depicted in Figure 2. Also, the refined stellar parameters of inclination, effective temperatures of primary and secondary components and luminosity of primary components were obtained from the DC2015 process, and they were tabulated in Table 1, comparing with the previous study done by Kurtz et al. (2015). Additionally, the values of newly introduced stellar parameters from the DC2015 modeling were also listed in Table 1.

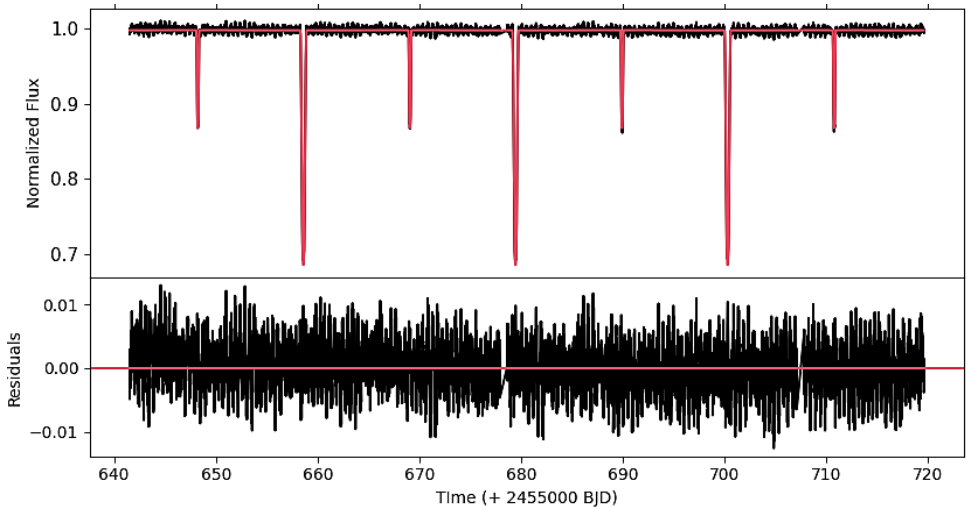

**Figure 1.** Upper panel: the results of light curve modeling of the KIC 8569819 system using WD DC2015 program. The Observed light curve (black line) for the system and theoretical DC2015 model (red line). Lower panel: the residuals of the best fitting model (black line) after subtraction of the DC models

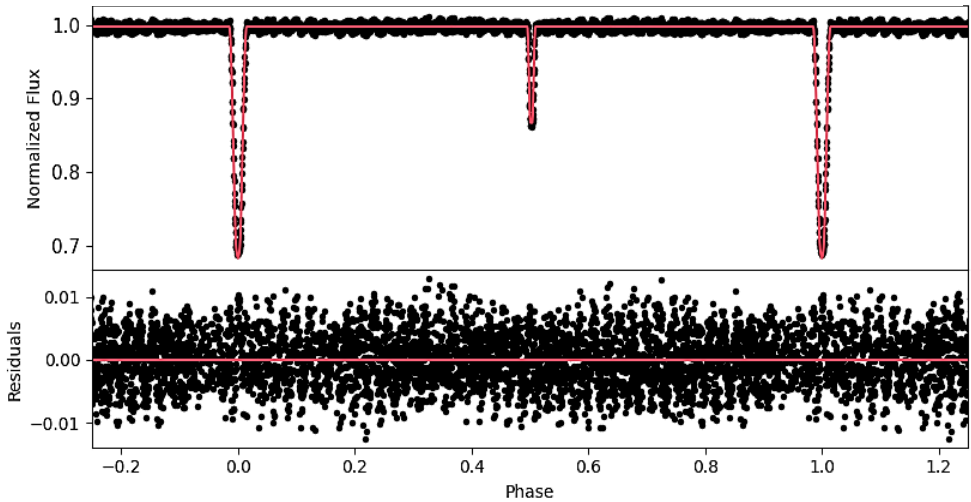

**Figure 2.** Upper panel: the phased light curve of the system (black points) with its final fit (red line). Lower panel: the residuals of the best fitting model (black points) after subtraction of the model

**Table 1.** Resulting output stellar parameters of KIC 8569819 generated from the WD DC2015 modelling

|  | Our study | | Kurtz et al. (2015) | |
|---|---|---|---|---|
|  | **Primary** | **Secondary** | **Primary** | **Secondary** |
| $T_{eff}$ (K) | 7155 $\pm$ 9 | 5956 $\pm$ 7 | 7100 $\pm$ 250 | 6047 $\pm$ 5 |
| M ($M_\odot$) | 1.724 | 1.014 | 1.7 | 1.0 |
| R ($R_\odot$) | 1.790 | 0.986 | - | - |
| log g (cm s$^{-2}$) | 4.17 | 4.46 | 4.0 | - |
| L ($L_\odot$) | 10.911 $\pm$ 0.005 | - | 10.958 $\pm$ 0.006 | - |
| $M_{bol}$ (mag) | 2.56 | 4.65 | 2.8 | - |
| log ($L/L_\odot$) | 0.877 | 0.040 | - | - |
| i (degrees) | 89.88 $\pm$ 0.03 | | 89.91 $\pm$ 0.04 | |

**SUMMARY, DISCUSSION AND CONCLUSION**

A comprehensive modeling of KIC 8569819 eclipsing binary system, which contains a δ Scuti component, was performed in this study to refine the stellar parameters and to introduce new stellar parameters of the primary and secondary components of the binary system. The LC2015 program and the DC2015 program exist in the WD modeling program, were applied to achieve the main goal of this study. The more precise application of the DC2015 process in the WD program was the newly added concept in this study compared to the previous modeling studies done for the KIC 8569819 eclipsing binary system. The results tabulated in Table 1 represent the summary chart for the findings of our study compared to the work done by Kurtz et al. (2015) for the KIC 8569819 target system. The application of this DC2015 process modifies the binary model as an iterative process by minimizing the residuals between observed and modeled light curves using the Levenberg-Marquardt algorithm concept. From that, we were able to extract the residual data, which is relevant to the δ Scuti type pulsating component in the system, by successfully removing the binary nature from the observed light curve. That extracted residual data can be used for frequency analysis in the field of asteroseismology to study the stellar interior of the δ Scuti type pulsating component. By enhancing the accuracy of the modeling of binary nature, we would be able to proceed precise frequency analysis as well. Rather than the precise light curve modeling for the KIC 8569819 target system in this study, we were able to introduce values for stellar parameters like radius (R), bolometric magnitude ($M_{bol}$) and surface gravity (log g), for primary and secondary stars in addition to the study of Kurtz et al. (2015). So, according to this study, we can conclude that the DC2015 in the WD program is a crucial step to successfully disentangle the binary characteristics from the observed light curves and to enhance the accuracy of the resultant stellar parameters.